\documentclass[preprint,showpacs,preprintnumbers,amsmath,amssymb]{revtex4}

\usepackage{graphicx}
\usepackage{dcolumn}
\usepackage{bm}

\begin{document}

\preprint{APS/123-QED}

\title{On the gap between an empirical distribution and 
an exponential distribution of waiting times 
for price changes in a financial market\\
}

\author{Naoya Sazuka}
\email{Naoya.Sazuka@jp.sony.com}
\affiliation{
Sony Corporation, 4-10-18 Takanawa Minato-ku, Tokyo, 108-0074 Japan\\
}

\date{\today}

\begin{abstract}
We analyze waiting times for price changes 
in a foreign currency exchange rate. 
Recent empirical studies of high frequency financial data support
that trades in financial markets do not follow a Poisson process 
and the waiting times between trades are not exponentially distributed. 
Here we show that our data is well approximated by a Weibull distribution 
rather than an exponential distribution in a non-asymptotic regime.
Moreover, we quantitatively evaluate how much an empirical data 
is far from an exponential distribution using a Weibull fit. 
Finally, we discuss a phase transition between a Weibull-law and a power-law 
in the asymptotic long waiting time regime.
\end{abstract}

\maketitle

\section{\label{sec:level1}INTRODUCTION}
It seems natural to assume that trades in financial markets arrive 
according to a Poisson process and 
the waiting times, which are the time intervals between trades, 
follow an exponential distribution~\cite{rf:1,rf:2}. 
However, on the other hand, 
recent empirical studies~\cite{rf:3,rf:4,rf:5,rf:6} 
observed that the waiting time distribution is non-exponential in different markets. 
Therefore, in order to understand market behavior quantitatively and systematically, 
it would be important to check validity of 
the exponential distribution hypothesis.

In this paper, 
we test validity of the exponential distribution hypothesis 
of waiting times for price changes using a real market data.
Then we evaluate a gap quantitatively between an empirical distribution 
of waiting times and an exponential distribution. 
In order to measure the gap, we introduce a more general distribution 
which includes an exponential distribution as a special case. 
It is desirable that the distribution can quantify
the gap using a small number of parameters. 
That is, by fitting such a distribution to a real market data, 
we would like to determine 
how much an empirical data is far from an exponential distribution 
through the parameters. 
For this purpose, one of the good example
is a Weibull distribution~\cite{rf:7}. 
A Weibull distribution is often used to model the time to failure 
and described by two parameters $m$ and $\alpha$ as follows. 
\begin{eqnarray}
\mbox{P.D.F.}\quad f(t)&=&\frac{m}{\alpha}\left(\frac{t}{\alpha}\right)^{m-1}
\exp{
\left[-\left(\frac{t}{\alpha}\right)^{m}\right]
}   
\quad 0 \le t < \infty, \alpha>0, m>0
\label{(1)}\\
\mbox{C.D.F}\quad F(t)&=&1-
\exp{
\left[-\left(\frac{t}{\alpha}\right)^{m}\right]
}   
\label{(2)}
\end{eqnarray}

\begin{figure}[ht]
\centerline{\includegraphics[width=9cm]{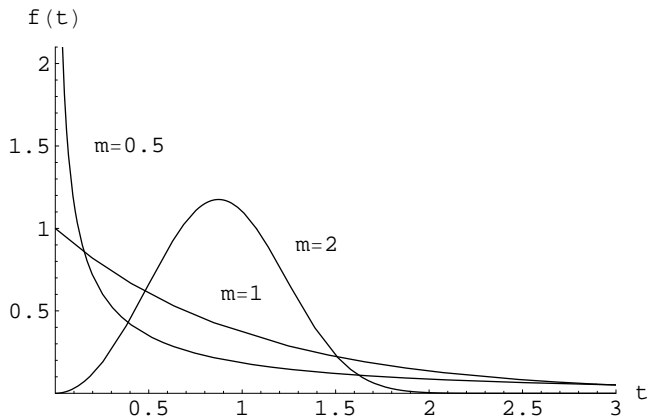}}
\caption{ Probability density functions of the Weibull distribution for different m's with $\alpha=1$}
\label{fig:1}
\end{figure}

where $\alpha$ is the scale parameter and $m$ is the shape parameter.
Since the distribution shape is characterized by only one parameter $m$,
we focus on the parameter $m$ in the following. 
FIG.~\ref{fig:1} shows some of probability density functions for different m's such as $m=0.5,1,2$. 
As is clear from FIG.~\ref{fig:1} and equation (\ref{(1)}), 
a Weibull distribution is reduced to an exponential distribution when $m=1$.  
Consequently, we can evaluate the gap by examining how much the estimated value $m$ is far from $1$.
Therefore, the purpose of this paper is to test validity 
of the exponential distribution hypothesis using a real market data 
and measure the gap quantitatively between an empirical distribution 
and an exponential distribution by fitting a Weibull distribution to the data.

It should be noted that we focus on a non-asymptotic regime 
for relatively short waiting times, since almost all events occur in this regime.
We discuss the asymptotic behavior for long time regimes in Section 4.

The paper is organized as follows. 
In Section 2, we explain our data and test the exponentianl hypothesis of waiting times.
In Section 3, by fitting a Weibull distribution to the empirical data,
we measure concretely the gap between the empirical data and an exponential distribution in two different ways.
In Section 4, we discuss the asymptotic behavior for long waiting time regime.
Finally, we present our conclusions in Section 5.

\section{Analysis of waiting times for price changes}
\subsection{Data}
In this paper, we analyze Sony bank USD/JPY rate as a real market data. 
Sony bank rate is that the Sony bank~\cite{rf:8} offers to their customers 
on their online foreign exchange trading service via the internet. 
The Sony bank rate depends on the market rate, but not customers' order. 
If the market rate changes by 0.1 yen and over, 
the Sony bank rate is updated to the market rate. 
Conversely, if the market changes by less than 0.1 yen, 
the Sony bank does not move and keep the previous rate. 
In other words, Sony bank rate is produced by filtering market rate
using the sort of {\it window} with the width of $\pm$0.1 yen.
In principle, the Sony bank rate is provided while the market is open. 
Currently, about 130,000 customers use this service.  
Our data set is about 31,000 data for the period of September 2002 to May 2004. 
According to the update rule, 
the mean time intervals between price changes of the Sony bank rate 
($\sim$20 minutes)~\cite{rf:9} is longer than the one of the market rate ($\sim$ 7 seconds). 

\subsection{Waiting times for price changes}
\begin{figure}[htb]
\begin{center}
\begin{tabular}{cc}
\resizebox{8cm}{!}{\includegraphics{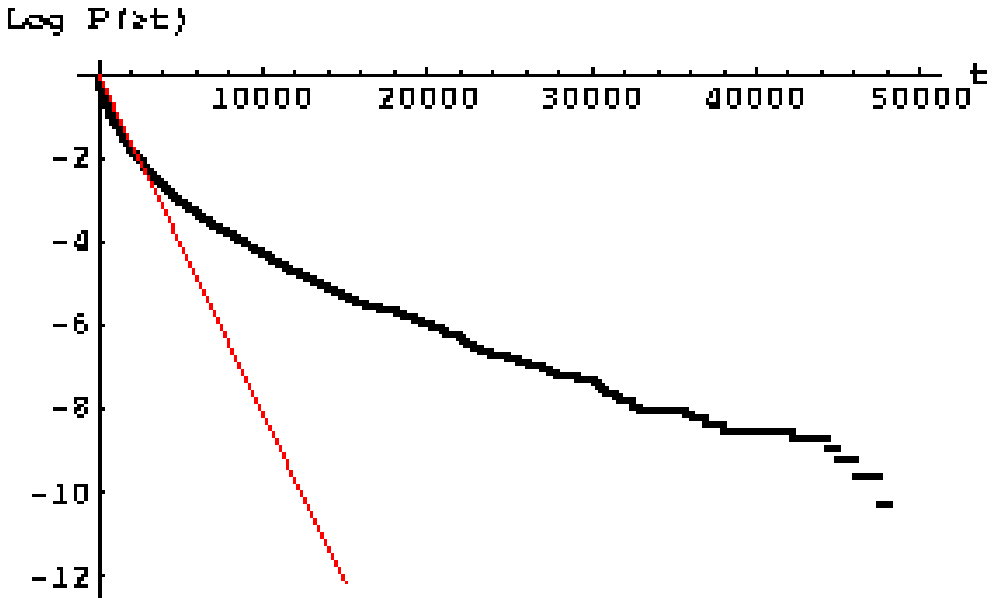}}
\resizebox{8cm}{!}{\includegraphics{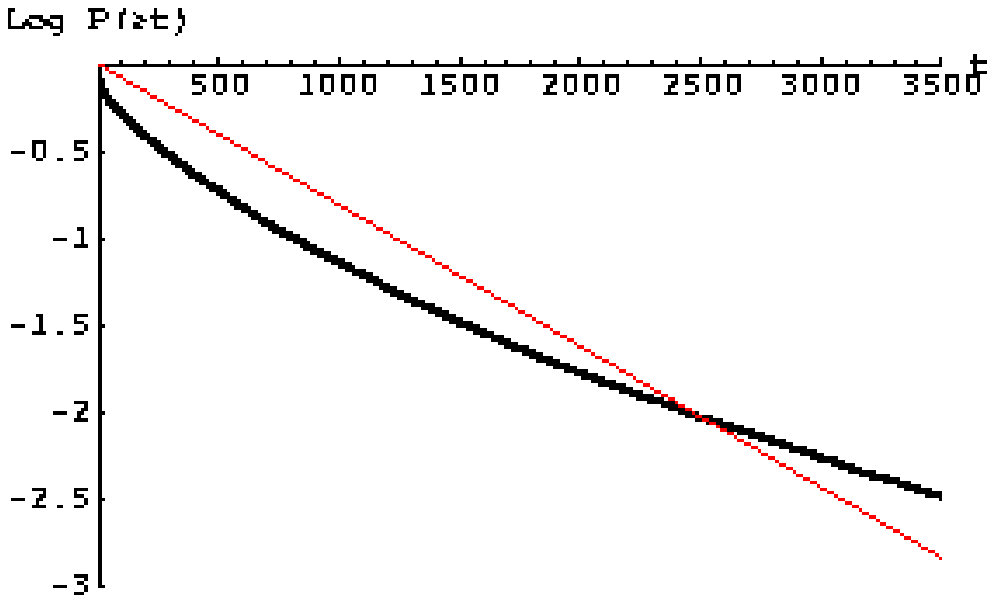}}
\end{tabular}
\end{center}
\caption{the semi-log plot of the survival function of waiting time $P(\ge t)$ 
and a fitted exponential distribution (a straight line).}
\label{fig:2}
\end{figure}

A waiting time $t_i$ between $i$th price change and $i+1$th price change 
is defined as follows.
\begin{equation}
t_i=s_{i+1}-s_{i}
\label{(3)}
\end{equation}
where $s_i$ is the time when $i$th price change occurs.
First of all, we plot a survival function of waiting time $P(\ge t)=1-F(t)$, 
which is the cumulative probability of the waiting times greater or equal to $t$ seconds, 
on a semi-log scale in FIG.~\ref{fig:2}. 
It shows that the waiting time for price changes is not exponentially distributed. 
If the distribution is exponential as is widely assumed, 
the data should be roughly on a straight line on the semi-log scale. 
However, we observe that the plotted data in FIG.~\ref{fig:2} is not a straight line.
The right panel of FIG.~\ref{fig:2} shows that
the gap 
is already visible in the very short waiting times regime.
This fact is consistent with recent empirical evidence observed in different markets~\cite{rf:3,rf:4,rf:5,rf:6}. 
Thus, we find that the non-exponential waiting time distribution appears 
not only in market rate but also in the data has been sampled by filtering market rate 
using the window with the width of $\pm$0.1 yen. 
The non-exponential waiting time distribution also means that 
the arrival process for price change is not a Poisson process. 

However, this semi-log plot of the cumulative probability $P(\ge t)$ is not enough to 
evaluate the gap quantitatively between an empirical distribution and an exponential distribution. 
As we explained in the previous section, 
we now quantify the gap by fitting a Webull distribution to the data. 
In particular, we check the gap visually by a Weibull paper analysis.

\section{The gap between empirical and exponentianl distriutions}
\subsection{Weibull paper analysis}
\begin{figure}[htb]
\centerline{\includegraphics[width=9cm]{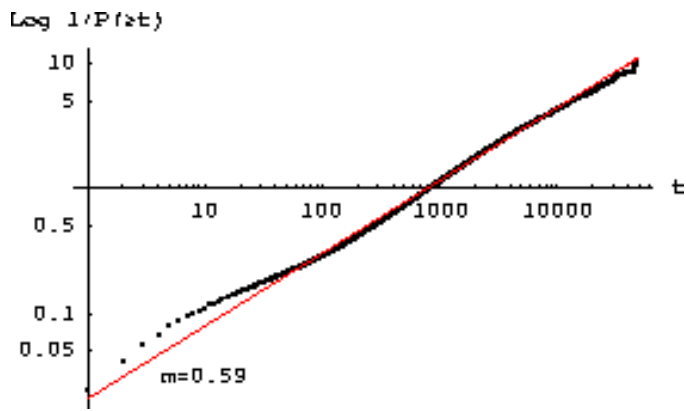}}
\caption{A Weibull plot of the Sony bank USD/JPY rate. 
The straight line is the estimated Weibull distribution with $m=0.59$.}
\label{fig:3}
\end{figure}

A Weibull paper analysis is often used to check a Weibull model assumption. 
A Weibull cumulative distribution function can be rewritten as 
\begin{equation}
\ln\ln\left(\frac{1}{1-F(t)}\right)
=m \ln t -m\ln\alpha .
\label{(4)}
\end{equation}
$Y=\ln\ln\left(1/\left(1-F(t)\right)\right)=\ln\ln(1/P(\ge t))$ 
is a linear function of $X=\ln t$ with a slope $m$. 
Namely, the data from a Weibull distribution 
are plotted on a straight line on Weibull paper. 
As a special case, the slope is 1 when the data follow an exponential distribution.
Therefore, the slope $m$ of the line on Weibull paper is a quantitative indicator
which enable us to evaluate the gap. 
In FIG.~\ref{fig:3}, our data is roughly on a straight line with the estimated slope $m\simeq0.59$,
which is apparently different from exponential distribution with $m=1$.
It shows that the waiting time distribution of the Sony bank USD/JPY rate 
is well approximated by a Weibull distribution with $m=0.59$,
rather than an exponential distribution.

\subsection{Divergence measurements}
The gap can be also discussed by using divergence measurements. 
In the previous section, we showed the empirical waiting time distribution
is approximated by a Weibull distribution.
Thus, we replaced 
the gap between an exponential distribution and the empirical data
with the gap between an exponential distribution and a Weibull distribution. 
However, in this section, we actually measure both of them
by considering a gap between two distributions as a divergence measurement.	
For example, we calculate Kullback-Leibler (KL) divergence and Hellinger distance.
These two divergence measurements between the empirical distribution $P$ 
and a model distribution $Q$ are written respectively
\begin{eqnarray}
\mbox{KL divergence}&=&\sum^{t_{max}}_{t=1}
P(t) \ln \left(P(t)/Q(t)\right)\\
\mbox{Helliger distance}&=&\sum^{t_{max}}_{t=1}
2\left(
P(t)-Q(t)
\right)^{2}.
\end{eqnarray}

\begin{table}[ht]
\begin{center}
\begin{tabular}{|l|c|c|}\hline
& KL divergence & Hellinger distance \\ \hline
Q=Weibull & 0.19 & 0.21 \\ \hline
Q=exponential & 0.49 & 0.36 \\ \hline
\end{tabular}
\end{center}
\caption{KL divergence and Hellinger distance between empirical distribution $P$ 
and a model distributon $Q$ $(t_{max}=50,000)$.}
\label{tab:1}
\end{table} 
TABLE~\ref{tab:1} gives KL divergence and Hellinger distance
when $Q$ is the fitted Weibull distribution with $m=0.59$ 
and the fitted exponential distribution.
Both divergence measurements show that the Weibull distribution is
closer than the exponential distribution to the empirical distribution. 
Especially, Hellinger distance which is a distance metric
shows that the difference is about 1.7 times. 
This fact correponds to the result in the previous section 
that the Weibull distribution is a better approximation 
than exponential distribution.

As a result, we conclude that the gap can be evaluated by 
two different methods which are a Weibull paper analysis and divergence measurements.
This is a main result of this paper.

\section{Phase transition between a Weibull-law and a power-law}

\begin{figure}[htb]
\begin{center}
\includegraphics[width=8cm]{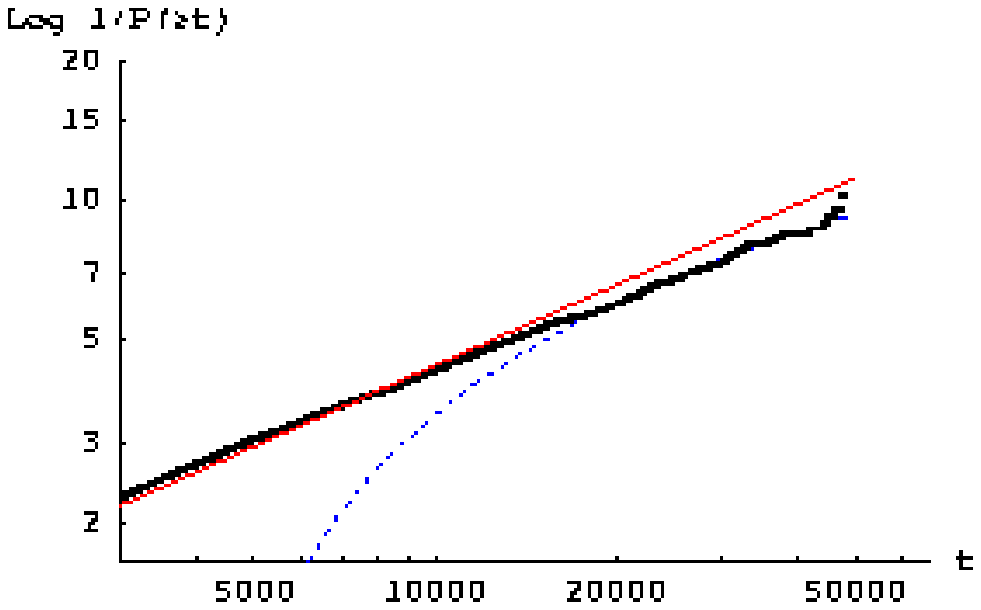}
\includegraphics[width=8cm]{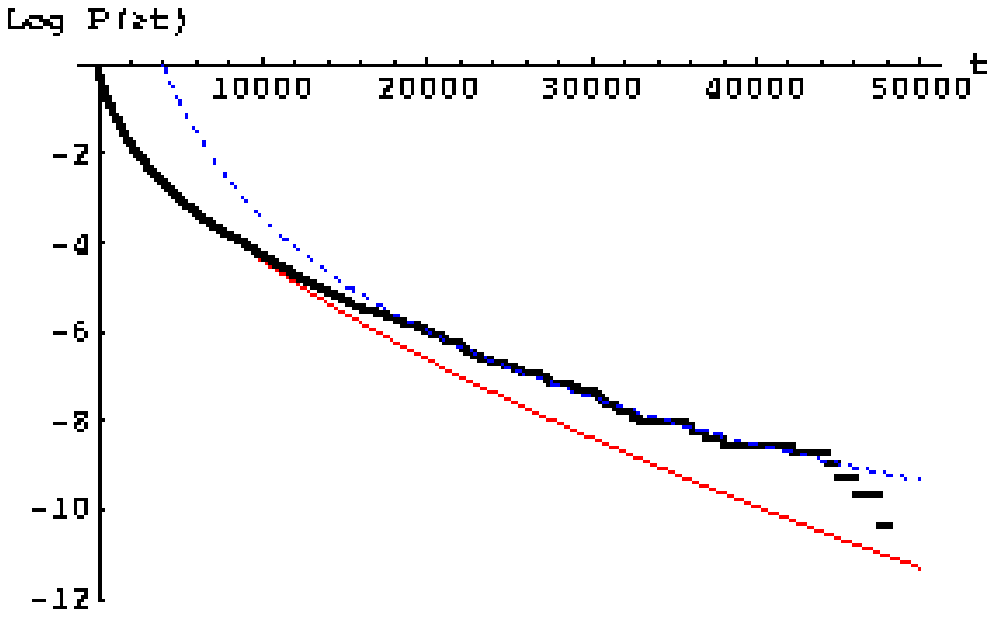}
\end{center}
\caption{The left panel is a Weibull paper in the asymptotic region and 
the right panel is the semi-log plot of the survival function of waiting time $P(\ge t)$.
Both shows the clear transition between a Weibull-law and a power-law. 
The thick line is the empirical data and 
the thin line is the Weibull distribution with m=0.59 and 
the dashed line is the power funciton with $\alpha$=3.67. }
\label{fig:4}
\end{figure}

Let us close this paper with a discussion on asymptotic behavior of waiting times.
In previous sections, we showed that a Weibull distribution is a better approximation 
of waiting times in a non-asymtiotic regime rather than an exponential distribution.
However, accurately the Weibull fit is 
not so good in the very short waiting time limit and the very long waiting time limit in FIG 3. 
It could have different behavior in those two time regimes. 
In this section, we discuss asymptotic behavior in long time limit which is one of two regimes. 
We find that the cumulative probability distribution $P(\ge t)$ of waiting times 
greater or equal to $\sim$18,000 seconds (= 5 hours) well expressed 
by a power-law $P(\ge t)\sim t^{-\alpha}$ with exponent $\alpha\sim 3.67$. 
Consequently, we find that the behavior changes from a Weibull-law to a power-law 
at some point $t_c\sim18,000$. 
The phase transition is well observed in the Weibull paper for the asymptotic regime 
and the semi-log scale of $P(\ge t)$ in FIG.~\ref{fig:4}. 
It should be noted that the number of data for the power-law regime is 
only 0.3$\%$, which is almost an outlier, of total number of waiting times. 
Since such long waiting times hardly ever happen, 
we focused on the Weibull fit regime in previous sections. 

Similar behavior was reported for the waiting time distribution of 
BUND futures traded at LIFFE~\cite{rf:6}. 
They showed that the waiting time distribution is 
a good agreement with the Mittag-Leffler function
which interpolates between
the stretched exponential 
$\exp\{-(\gamma t)^{\beta}/\Gamma(1+\beta)\}$
for short time regime 
to the power-law 
$(\gamma t)^{-\beta}/\Gamma(1-\beta)$
for long time.
It is interesting that even the Sony bank rate which is filtered market rate 
firstly moves out from a range of $\pm$0.1 yen  
still have similar property to the market rate itself.
However, the exponent values and transition point are different from our case. 
Our exponent value changes from $m\simeq0.59$ for a Weibull-law regime 
to $\alpha\simeq3.67$ for a power-law regime, 
whereas, the Mittag-leffler fit for BUND futures can be expressed by 
a single exponent $\beta\simeq0.95$ for both regimes. 
Moreover, our exponent value $m\simeq0.59$ for non-asymptotic regime is 
smaller than $\beta\simeq0.95$ for BUND data
and our transition point $t_c\sim$ 18,000 seconds is larger than 
$t_c\sim100$ seconds for BUND futures. 
This could be caused by the window effect 
which is the sampling method of the Sony bank rate from the market rate. 
By this sampling method, the waiting times of the Sony bank rate become longer 
than the one of the market rate on average.
The cumulative distribution function of the Weibull distribution with $m<1$ 
decays slower for $t>\alpha$ as $m$ decreases, in other words,
there are more long waiting times for $t>\alpha$ as $m$ decreases.

On the other hand, the behavior differs from a Weibull fit
in short time limit is not resolved. 
One possible reason is measurement errors but 
there might be other possible distributions that decay faster than the Weibull distribution.
The Mittag-Leffler fit also can not capture the empirical data for short time limit~\cite{rf:6}.

\section{CONCLUSION}
In this paper, we showed that the waiting time distribution of 
the Sony bank USD/JPY rate is non-exponential. 
The result is consistent with the recent empirical evidence~\cite{rf:3,rf:4,rf:5,rf:6} of market rate.
It is interesting that the non-exponential waiting time distribution appears 
not only in the market rate 
but also in the Sony bank rate which is filtered market rate
using the window with the width of $\pm$0.1 yen.

We also showed that the empirical data is well approximated by a Weibull distribution
in a non-asymptotic regime, which includes almost all events. 
Then we measured the gap quantitatively between an empirical waiting time distribution 
and an exponential distribution by using a Weibull paper analysis and divergence measurements.
Finally, we found that the phase transition between a Weibull-law and a power-law
in long time asymptotic regime.
It should be noted that the events in a power-law regime 
is only 0.3$\%$ of the total events.

In addition, the window effect of the Sony bank rate is can be regarded as 
the first passage time problem.
Further analysis of waiting times in this direction
will be reported in our forthcoming article~\cite{rf:10}.

\begin{acknowledgments}
I would like to appreciate Shigeru Ishi, President of the Sony bank, 
for kindly providing the Sony bank data and useful discussions.
Stimulating discussions with Jun-ichi Inoue of Hokkaido University is acknowledged.
\end{acknowledgments}

\bibliography{apssamp}

\end{document}